\documentstyle[aps,twocolumn]{revtex}

\begin{document} 
\draft

\title{Exact Ground State Properties of Disordered Ising-Systems}
\author{J.~Esser, U.~Nowak, and K.~D.~Usadel} 
\address{ 
  Theoretische Tieftemperaturphysik\\ 
  Gerhard-Mercator-Universit\"{a}t-Duisburg\\
  47048 Duisburg/ Germany\\
  e-mail: joerg@thp.uni-duisburg.de 
} 

\date{\today}
\maketitle

\begin{abstract}
  Exact ground states are calculated with an integer optimization
  algorithm for two and three dimensional site-diluted Ising
  antiferromagnets in a field (DAFF) and random field Ising
  ferromagnets (RFIM), the latter with Gaussian- and
  bimodal-distributed random fields. We investigate the structure and
  the size-distribution of the domains of the ground state and compare
  it to earlier results from Monte Carlo simulations for finite
  temperature.

  Although DAFF and RFIM are thought to be in the same universality
  class we found essential differences between these systems as far as
  the domain properties are concerned.  For the DAFF the ground states
  consist of fractal domains with a broad size distribution that can
  be described by a power law with exponential cut-off.  For the RFIM
  the limiting case of the size distribution and structure of the
  domains for strong random fields is the size distribution and
  structure of the clusters of the percolation problem with a field
  dependent lower cut-off. Consequently, the domains are fractal and
  in three dimensions nearly all spins belong to two large infinite
  domains of up- and down spins - the system is in a two-domain state.
  The fractal dimensions for the DAFF and the RFIM agree. The DAFF
  ground state properties agree with results from MC simulation in the
  whole whereas there are essential differences between our exact
  ground states calculations and earlier MC simulations for the RFIM.
\end{abstract}

\pacs{PACS: 75.50.Lk, 64.60.Cn, 75.40Mg\\
  Keywords: Ising-Models, Random Magnets, Numerical Methods}

\section{introduction}
Many aspects of the influence of random field disorder on a system of
interacting spins are still not well understood (for a review see
\cite{belanger}).  Since it has been argued that the diluted Ising
antiferromagnet in a homogenous field (DAFF) and the random field
Ising ferromagnet (RFIM) are in the same universality class
\cite{fishman,cardy} experimental investigations focus on the DAFF
\cite{kleemann} as an experimental realizations of a system with
random field disorder while theorists usually focus on the RFIM
\cite{nattermann}.

However, there have been investigations on the domain structures of
the RFIM \cite{cambier} and the DAFF \cite{nowak1,nowak2}, both based
on MC simulations which suggest that there might be essential
differences between DAFF and RFIM, at least as far as the domain
structure is concerned. In the limit of strong disorder for the RFIM
the domains are fractal in three and compact in two dimensions while
the domains of the DAFF have a fractal structure both in two and three
dimensions. Additionally, due to the fractality of the domains
essential deviations from Imry-Ma-type \cite{imry} behavior which is
thought to be valid for small disorder have been found.

In order to prove how far the results from MC simulations which are
out of equilibrium for these systems with frozen dynamics
\cite{villain} can be transferred to the equilibrium $T=0$ case we
performed exact ground state calculations on DAFF and RFIM with an
integer optimization algorithm. With this method we get exact
information on equilibrium properties at zero temperature. We
investigate the distribution of domain sizes and the structure of the
domains and we compare our findings for the DAFF and the RFIM.
Hopefully, the results may help to understand the critical behavior of
these systems as well as dynamical aspects \cite{nowak3}.

The Hamiltonian of the DAFF in units of the nearest neighbor coupling
constant $J$ is
\begin{equation}
  H=\sum_{<ij>}\epsilon_i \epsilon_j \sigma_i \sigma_j - B\sum_i \epsilon_i \sigma_i
\end{equation}
with the uniform field $B>0$ on all sites of the quadratic $L \times
L$ respectively cubic $L \times L \times L$ lattice. Here
$\sigma_i=\pm 1$ denotes Ising-spins and a fraction $p$ of the sites
is occupied with a spin (quenched disorder: $\epsilon_i = 0,1$). The
phase diagram of the 2d-DAFF consists of an antiferromagnetic low
temperature phase for magnetic field $B=0$ and a disordered phase for
all finite values of $B$. In three dimensions there exists a
long-range ordered phase \cite{imbrie,bricmont} for magnetic fields
$B$ smaller than the critical fields $B_c$ which is $B_c \approx 1.4$
for a dilution of $p = 0.5$ \cite{nowak1}. For higher fields and low
temperatures the DAFF develops a frozen disordered domain state in two
and three dimensions. This domain state has many of the
characteristics of a spin glass, as for instance a remanent
magnetisation and an irreversibility line scaling like the
deAlmeida-Thouless line \cite{nowak1}.

The Hamiltonian of the RFIM in units of the nearest neighbor coupling
constant $J$ is 
\begin{equation}
H=-\sum_{<ij>} \sigma_i \sigma_j - \sum_i B_i \sigma_i.
\end{equation}
Here, all sites are occupied and the random fields $B_i$ are taken from a
Gaussian (Gaussian-RFIM) respectively bimodal ($\pm\Delta$-RFIM) probability
distribution corresponding to
\begin{equation}
P(B_i)=\frac{1}{\sqrt{2\pi}\Delta} e^{-(B_i/\sqrt{2}\Delta)^2}
\end{equation}
and
\begin{equation}
P(B_i)=\frac{1}{2}(\delta(B_i-\Delta)+\delta(B_i+\Delta)),
\end{equation}
respectively.

As in the case of the DAFF the phase diagram of the 2d system consists
of a long-range ordered low temperature phase for zero random fields
and a disordered phase for all finite values of the random field
\cite{imbrie}.  In three dimensions for weak random fields and low
temperatures there exists a long-range ordered phase \cite{bricmont}.
The critical value for the random field at which long-range order
disappears was found to be $\Delta_c = 2.35$ at zero temperature
\cite{ogielski}. Even for the RFIM a possible glassy transition has
been discussed \cite{mezard}.

We calculate exact ground states using an optimization algorithm well
known in graph-theory. The Ising-system is mapped on an equivalent
transport network, and the maximum flow is calculated using the
Ford-Fulkerson algorithm \cite{picard,ford,hartmann}. In order to
investigate the distribution and the structure of domains we perform a
cluster analysis with a suitable adjusted Hoshen-Kopelman type
algorithm \cite{hoshen}. This algorithm pieces the system into
domains, where a domain is defined as a group of spins which are
connected and antiferromagnetically (DAFF) respectively
ferromagnetically (RFIM) ordered.

The investigations for the DAFF are carried out for systems of size
$400 \times 400$ with an average over 50 different dilution
configurations with $p = 0.7$ for each value of $B$. For some
structural aspects even ground states of size up to $700 \times 700$
were calculated. Systems of size $50 \times 50 \times 50$ averaged
over 40 dilution configurations with $p=0.5$ are used for $d=3$.  The
calculation of ground states for the RFIM is more time consuming.
Here we considered $300 \times 300$-systems with 50 different random
field configurations in 2d and 40 systems of size $30 \times 30 \times
30$ in 3d.

\section{DISTRIBUTION OF DOMAIN-SIZES}
\subsection{DAFF}
In two dimensions for $B>0$ no long-range order exists and the ground
state of the system consists of antiferromagnetic domains of finite
size.  Within a ground state for each domain the domain-wall energy
$E_w$, i.~e.~the number of broken bonds, must be smaller than the
field-energy $B m_D$, where $m_D$ is the absolute value of the
magnetisation of the domain which is not zero in these
antiferromagnetic domains due to the dilution. Therefore, only domains
with $m_D \geq E_w / |B|$ exist in the system. The minimum wall
energy of a domain is one broken bond, $E_w = 1$.  Additionally, it is
always $V_D \geq m_D$, where $V_D$ is the size of the domain (number
of spins), so that with $V_D \geq 1/|B|$ we get a rigorous lower cut-off
for possible domain sizes.  This relation indicates that except from
the isolated spin-clusters only large domains can be realized for very
weak fields. Due to this finite size effect we consider here fields $B
\geq 0.5$ only and hence, all domain sizes $V_D \geq 3$ exist.

As has been shown earlier \cite{nowak3} for the two dimensional case
the distribution of domains in a DAFF is well described by a power law
with an exponential cut-off,
\begin{equation}
  N_D(V_D) \sim V_D^{-\delta}\exp (-V_D/V_0).
\label{nv}
\end{equation}
where $V_0$ depends strongly on the field. Because of the inaccuracy
in the determination of $V_0$ no significant $p$-dependence can be
observed for this quantity. We increased the numerical effort in order
to investigate a possible concentration and field dependence of
$\delta$. We find $\delta$ weakly decreasing for increasing field but
no $p$-dependence. The results are summarized in Tab.~\ref{deltatable}. 

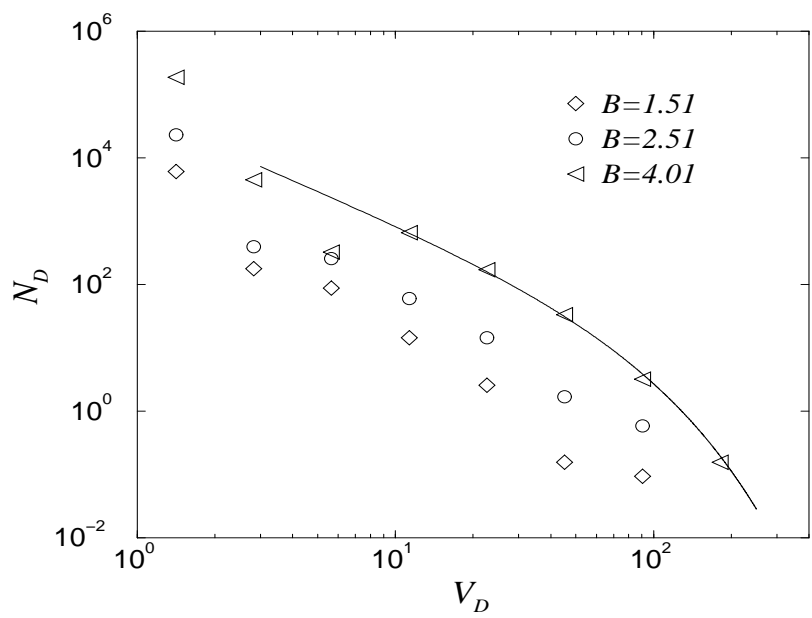
\begin{figure}
  \caption{$N_D$ versus $V_D$ for the DAFF in three dimensions. For
    $B=4.01$ a fit to Eq.~\ref{nv} is also shown as solid line.}
  \label{abbruchdaff3d}
\end{figure}

Fig.~\ref{abbruchdaff3d} shows the size-distribution of the domain
state of the three dimensional DAFF for different fields.  For the
largest value of the field a reasonable fit to Eq.~\ref{nv} can be
carried out yielding $\delta = 1.8 \pm 0.3$ and $V_0=50\pm 20$.
However, a detailed investigation of the field or concentration
dependence of $V_0$ and $\delta$ is not possible with satisfactory
precision for the following reasons: As has been shown earlier
\cite{nowak2} for $B \approx B_c$ the domain state consists mainly of
two large interpenetrating domains (for $B=4$ these domains contain
approximately 80\% of the spins). In Fig.~\ref{abbruchdaff3d} the two
isolated peaks in the size distribution corresponding to the two
percolating domains are not shown but due to the existence of these
two infinite domains there are only few data for domains of finite
size so that the statistics for these domains is rather bad. In the
limit of very high fields the domains become very small. For $B > 6$
all spins are polarized by the field, i. e. all domains have size 1.
Hence, there is only a small region of values of the magnetic field
where a broad distribution of domain sizes exist. As we will show in
the next subsection the existence of two infinite domains has even
more dramatic consequences for the RFIM. Here, there are no values of
random fields where relevant domains of finite size can exist.

\subsection{RFIM}
In the limit of strong random fields $\Delta > 2d$ ($\pm \Delta$-RFIM)
respectively $\Delta \rightarrow \infty$ (Gauss-RFIM) all spins follow
the direction of the random field. Hence, the ferromagnetic domains
are the clusters of sites with uniform random field signs. In this
limit the domains of the RFIM correspond to clusters of the
percolation problem \cite{stauffer} with a concentration of $50\%$.

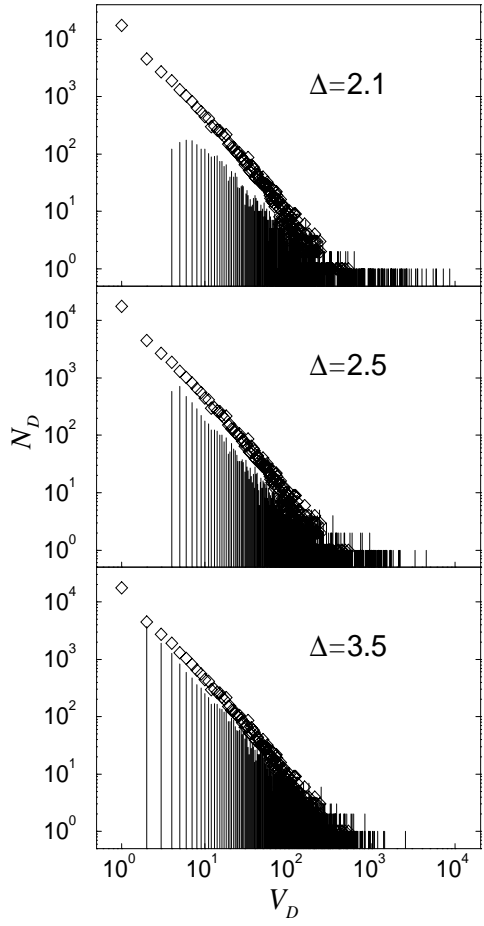
\begin{figure}
  \caption{Domain-size distributions for the two dimensional $\pm
    \Delta$-RFIM (Impulses) and size-distribution of the random field
    clusters (open symbols)}
  \label{nvpm2d}
\end{figure}

The distribution of domain-sizes for different $\Delta$ is shown in
Fig.~\ref{nvpm2d} for the two dimensional $\pm \Delta$-RFIM. For
comparison the size-distribution for the percolation problem with a
concentration of $50\%$ is also shown. With increasing $\Delta$, the
domain-size distribution approaches the cluster-size distribution of
the percolation problem which follows Eq.~\ref{nv} with $\delta = 1.55
\pm 0.05$ and $V_0=120\pm 20$.

For lower random fields there is a striking difference between the
Gaussian-RFIM on the one hand and the $\pm \Delta$-RFIM on the other
hand: for the latter system there is a lower cut-off for the possible
domain-sizes depending on the strength of the random field $\Delta$.
This cut-off can be seen in Fig.~\ref{nvpm2d} and it can be determined
analytically: Similar to the arguments given for the DAFF for each
domain the relation $E_w < E_B = \sum_{i=1}^V \sigma_i B_i$ must hold.
The positive domain-wall energy has to be compensated by the negative
field energy. In the most favorable case all random fields acting on
the spins of a domain have the same sign, $E_B = V_D \Delta$ so that
for the smallest possible domains we get $V_D \geq E_w/\Delta$. With
this relation as starting point a lower cut-off can be determined as
the sizes of those domains which have a minimum surface.  For large
domain sizes the minimum surface of a domain is the surface of a
circle (2d), $E_w \sim V_D^{1/2}$, respectively sphere (3d), $E_w \sim
V_D^{2/3}$. For smaller sizes, the domain with the minimum surface
$V_{min}$ can be determined numerically.

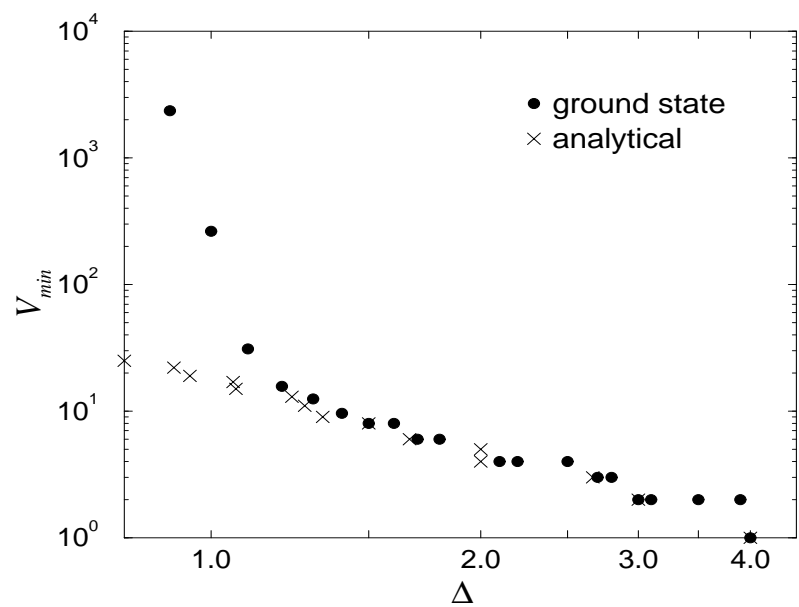
\begin{figure}
  \caption{Minimum domain size for given values $\Delta$ of the field
    of a $\pm \Delta$-RFIM in two dimensions}
  \label{minvpm2d}
\end{figure}

Fig.~\ref{minvpm2d} shows this dependence of $V_{min}$ from $\Delta$
for the $\pm \Delta$-RFIM in two dimensions. In addition, the minimum
domain-volumes of the ground states of the RFIM for different $\Delta$
are depicted.  For $\Delta \leq 1$ there is a strong deviation of the
observed domains of minimum size from the theoretical curve.
Obviously, this is due to the fact that for larger domains it is
less probable that all random fields within a domain with minimized 
surface have the same orientation. Hence, either the field energy $E_B$
is overestimated or the shapes of the domains are more complicated
than those with minimum surface. The first possibility would correspond to
a crossover to a behavior following the Imry-Ma-argument
\cite{imry},  where $E_B$ is thought to scale with the root of the
domain size due to simple statistical fluctuations of random fields
within a domain. However, in the range of fields where this crossover
may occur the minimum domain size is larger than the system sizes that
we can investigate with numerical methods. 

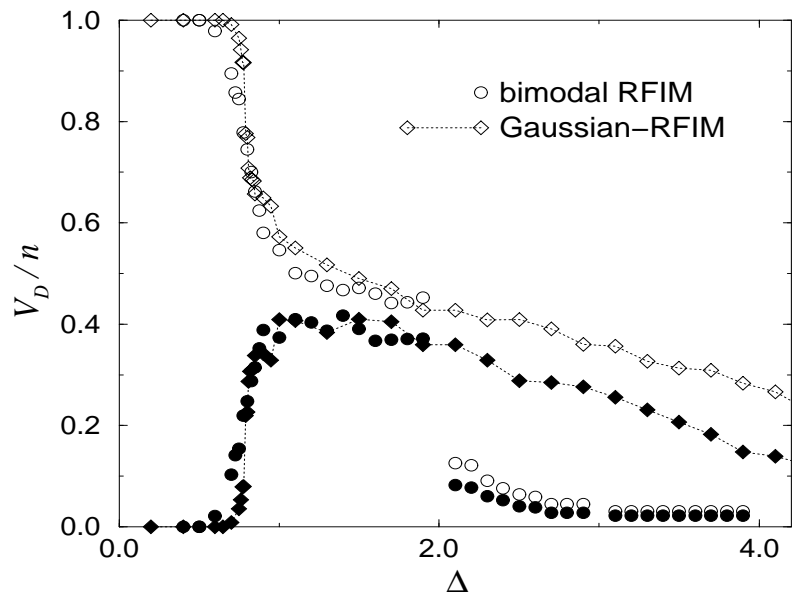
\begin{figure}
  \caption{Largest (open symbols) and second largest (filled symbols)
    domain size divided by the system size $L \times L$ for the $\pm
    \Delta$-RFIM in two dimensions}
  \label{maxpmgau2d}
\end{figure}

Fig.~\ref{maxpmgau2d} shows the volumes of the two largest domains in
the ground state of the $\pm \Delta$- and the Gaussian-RFIM. For lower
fields $\Delta < 0.8$ the order parameter is finite. Obviously, this
is the region where finite size effects lead to wrong results for the
system sizes we investigated since in this regime minimum domain
sizes are larger than the system.  Therefore, in the following we
consider only results for $\Delta > 0.8$.

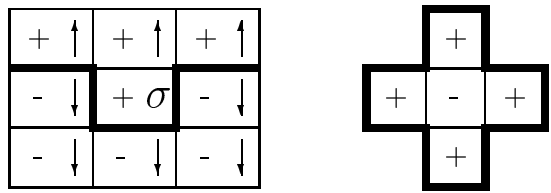
\begin{figure}
  \caption{$\pm \Delta$-RFIM; left: cut of a domain wall; right: a
    domain for $\Delta>2$ consisting of multiple random field clusters}
  \label{domainwall}
\end{figure}

An additional, striking feature of Fig.~\ref{maxpmgau2d} is the jump
in the data for the $\pm \Delta$-RFIM at $\Delta = 2$. Considering
Fig.~\ref{domainwall} (left) one realizes $\Delta = 2$ as a special
value: The random fields are marked with $+$ and $-$ signs and the
arrows represent the spin-orientations ($\uparrow$-spins are oriented
parallel to a positive $\Delta$).  The energy of the center spin with
$\sigma = \pm 1$ is
\begin{equation}
  E_{\sigma} = 2\sigma - \sigma \Delta.
\end{equation}
In the ground state it is $\sigma = +1$ for $\Delta > 2$ and the bold
drawn domain wall is realized. $\Delta < 2$ favors $\sigma = -1$ and
the center spin belongs to the down-domain. Generally, for $\Delta >
2$ the domain wall arranges in such a way that the sign of the
random fields changes at each position of a domain wall, i. e. the
domain walls run along the surfaces of the random field clusters
(clusters of sites with uniform random field sign).
Fig.~\ref{domainwall} (right) demonstrates that nevertheless even for
$\Delta > 2$ domains larger than these random field clusters can
exist. A cluster with a magnetisation that is oriented antiparallel to
the random fields is enclosed such that its surface lies within the
domain. Hence, for $\Delta > 4$ each domain is a random field cluster
while for $2 < \Delta < 4$ each domain wall is a part of the surfaces
of the random field clusters. Consequently, the domains in this range
of $\Delta$ are larger.  In three dimensions each spin has six next
neighbors. Here, the corresponding effect occurs at $\Delta = 4$.

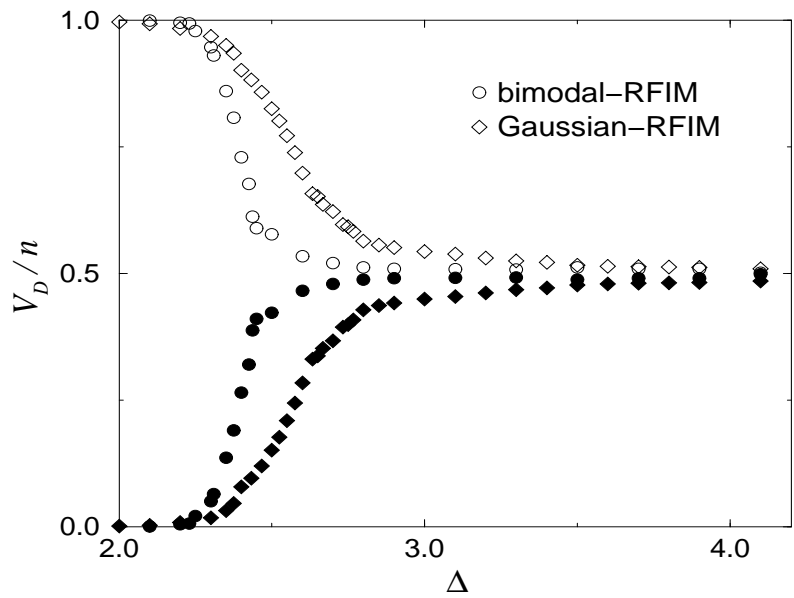
\begin{figure}
  \caption{Largest and second-largest domain size corresponding to
    Fig.~\ref{maxpmgau2d} for the RFIM in three
    dimensions.}
  \label{maxpmgau3d}
\end{figure}

Both, for $d=2$ and $d=3$ the RFIM approaches the percolation problem
with a concentration of 50\% for large $\Delta$, but while for $d=2$
in this case there exists no percolating cluster of random fields in
three dimensions 50\% is above the percolation threshold.  This means
that as mentioned above for the case of the DAFF in three dimensions
one will always find two infinite interpenetrating domains within the
disordered phase whose antiparallel oriented magnetisations cause the
order-parameter to vanish. The normalized volumes of the largest 
and second-largest domain for the $d=3$ $\pm \Delta$-RFIM are
shown in Fig.~\ref{maxpmgau3d}. Adding the sizes of these two
largest domains we found that more than $97\%$ of the spins of the
system belong to these two domains in the whole range of random
fields. In contrast to $d=2$ there is no broad range of domain-sizes
for $d=3$ and in contrast to the DAFF there is no range of fields
where a relevant number of domains of finite size exist. For this
reason it is senseless to analyze the distribution of the domains in
more detail.

\section{DOMAIN STRUCTURES}
\subsection{DAFF}
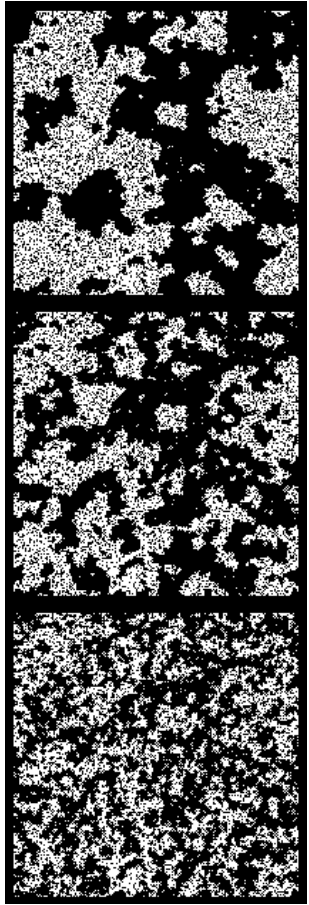
\begin{figure}
  \caption{Ground states of a $200 \times 200$ DAFF with a
    fixed dilution configuration for increasing fields $B = 0.8, 1.4,
    2.2$ (from above).  The two antiferromagnetic phases are
    represented black and white the vacancies are also black.}
  \label{daff2dfolge}
\end{figure}

In a DAFF domains are composed in such a way that as many spins as
possible are oriented parallel to the applied field with a minimum
number of broken bonds. Hence, domain walls favorably run along
vacancies to minimize the wall energy. These two competing
requirements for minimizing the total energy lead to highly
non-trivial structures.  Fig.~\ref{daff2dfolge} shows the
antiferromagnetic domains in the ground state. Here, the configuration
of occupied sites and vacancies is kept fixed, while the external
field $B$ is tuned.  In order to investigate the structural properties
of the domains we consider the following quantities and corresponding
scaling relations:
\begin{itemize}
\item volume $V_D$: number of spins
\item surface $S_D$: sum of all unsatisfied bonds (broken or "open"
  due to vacancies) with $S_D \sim V_D^{1/D_{vf}}$
\item radius of gyration $R_D$, i. e. root of mean squared distance
  of all spins of a domain with $ V_D \sim R_D^{D_v}$
\item absolute value of the magnetisation of a domain $m_D$ with $m_D
  \sim V_D^{\theta}$
\item wall energy $E_W$, sum of all broken bonds with $E_W \sim V_D^{\sigma}$
\end{itemize}
For large domains the scaling relations above hold and the exponents
are determined by analyzing all domains in the ground states and
taking an average of the quantities above for all domains of equal
size $V_D$. In \cite{nowak3} we showed an example for the two
dimensional case. In the meantime we investigated the three
dimensional system also. Our resulting exponents are summarized in
Tab.~\ref{daffstructuretable} and compared to the values resulting
from MC simulations \cite{nowak1,nowak2}.

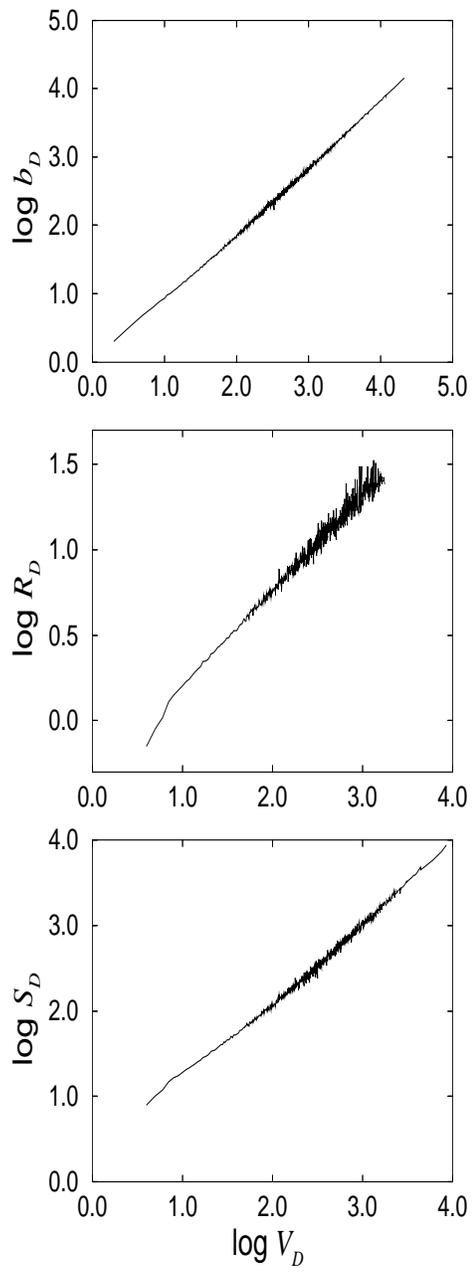
\begin{figure}
  \caption{Figures for the determination of the structural exponents $\theta$,
    $D_v$ and $D_{vf}$ for the two dimensional $\pm \Delta$-RFIM.}
  \label{exponentenpm2d}
\end{figure}

The domains are fractal with fractal dimension $D_v=1.65$ for $d=2$
and $D_v=2.14$ for $d=3$. This fractality is also reflected in the
proportionality of surface and volume ($D_{vf} \approx 1$). A great
part of the surface is inside the domains. Slight deviations of
exponents from exact ground states from those resulting from MC
simulation are due to a systematic inaccuracy which is not reflected
by the error-bars representing the fluctuations of the data only.
Probably, a small, systematic error is due to the fact that the data
are slightly curved in a log-log plot so that the linear range for
large domains can hardly be determined (see also the corresponding
Fig.~\ref{exponentenpm2d} for the RFIM).  From comparison with the
corresponding Figures \cite{nowak1,nowak2} it is found that in spite
of similar exponents the values $m_d(V_D)$ are larger for exact ground
states - the domains of the ground states are obviously better
optimized.

In good agreement with MC simulations is $\theta \approx 1$ both in
two and three dimensions. Altogether the fractal structures for
strong disorder lead to deviations from Imry-Ma-type arguments
\cite{imry} based on compact domains. Here, a statistically
distribution of vacancies in a domain would yield $\theta \approx 1/2$
which can clearly be ruled out by our data for the range of fields
that we investigated. A field dependence of any of the exponents could
not be found.

\subsection{RFIM}
As we showed in the previous section there is no broad distribution of
domains for the three dimensional RFIM - domains of finite size are
irrelevant for the ground state. Therefore, in this subsection we
restrict ourselves to the case $d=2$.

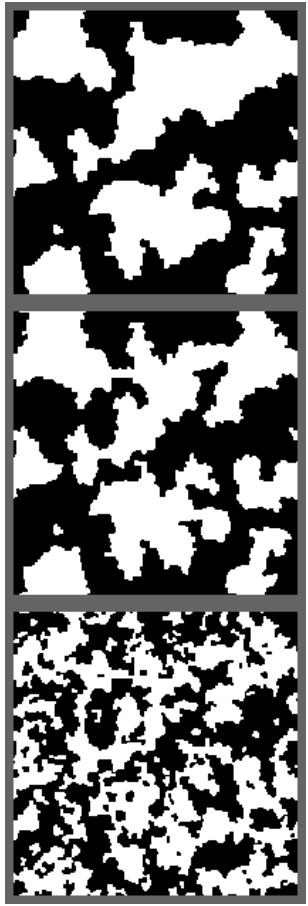
\begin{figure}
  \caption{Ground states of a $200\times 200$ $\pm \Delta$-RFIM for a
    fixed configuration of random fields and $\Delta = 1.0, 1.1, 1.5 $
    (from above). The two ferromagnetic phases are represented black
    and white.}
  \label{pm2dfolge}
\end{figure}

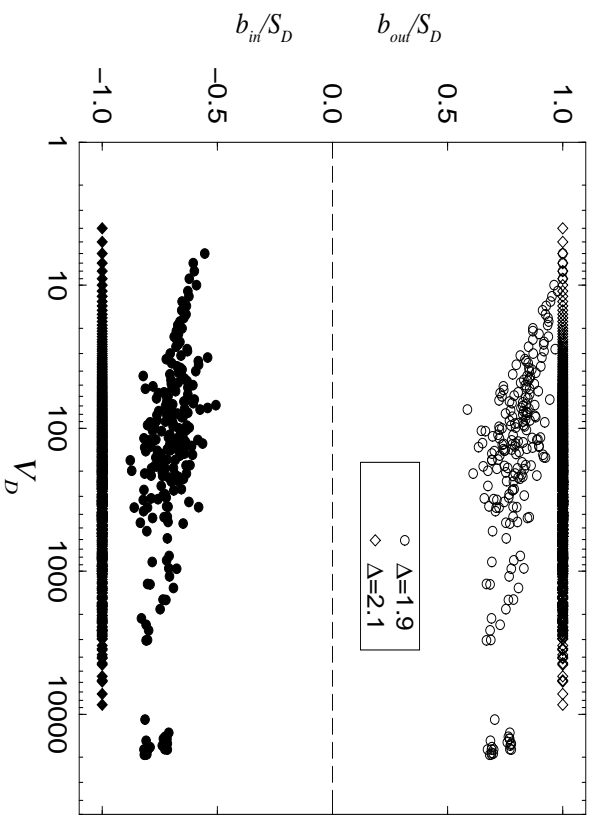
\begin{figure}
  \caption{Random field fluctuations $b_{out}$ and $b_{in}$ of the
    domain surfaces each multiplied with the sign of the
    domain-magnetisation and divided by the surface $S_D$ for the
    two dimensional $\pm \Delta$-RFIM.}
  \label{iopm2d}
\end{figure}

Fig.~\ref{pm2dfolge} shows the ferromagnetic domain configuration of a
$\pm\Delta$-RFIM in the ground state. The configuration of signs of
the random fields is kept fixed, while the value $\Delta$ of the field
is tuned. As for the DAFF we study the structural quantities volume,
surface and radius of the domains.  Since there is no dilution the
surface and the domain wall energy are identical. Also, the
magnetisation and the volume of the domains are identical. We study
the following quantities:
\begin{itemize}
\item local random field fluctuations $b_D$, which is the absolute
  value of the sum of the random fields of a domain
\item random field fluctuations of all sites of a domain along the
  domain wall inside the domain $b_{in}$
\item random field fluctuations of all sites of a domain along the
  domain wall outside the domain $b_{out}$
\end{itemize}
Fig.~\ref{iopm2d} shows $b_{in}/S_D$ and $b_{out}/S_D$ each multiplied
with the sign of the magnetisation of the considered domain for the
$\pm\Delta$-RFIM. As we demonstrated with Fig.~\ref{domainwall} for
$\Delta > 2$ the domain walls run exactly along the boundaries of
random field clusters. Hence, in this range of fields it is
$|b_{in}/S_D| = |b_{out}/S_D| = 1$. Even for smaller fields the domain
walls are presumably located along these random field surfaces and  
for large domains these fluctuations are proportional to the surface of
the domains.

Corresponding to the DAFF the exponents $D_v$, $D_{vf}$ and $\theta$
with $ b_D \sim V_D^{\theta}$ are determined from appropriate fits
(Fig.~\ref{exponentenpm2d}). In Tab.~\ref{rfimstructuretable} the
exponents are summarized. The exponents resulting from MC are from
\cite{cambier}.

The investigations are carried out for relative strong random fields,
$\Delta = 3.01$ (Gaussian-RFIM) and $\Delta = 4.01$ ($\pm
\Delta$-RFIM), because then - as explained in the previous chapter - a
broad range of domain sizes exists.  It is found that the exponents do
not depend on the kind of random field distributions within the given
precision.  

In agreement with the findings for the DAFF the domains are fractal
with nearly the same exponents $D_v \approx 1.6$ and $D_{vf} \approx
1$.  Also, it is $\theta \approx 1$ indicating that the
random fields are distributed non-statistically within a domain. These
results are in contradiction to the earlier investigations using MC
methods \cite{cambier} where results were found corresponding to more
compact domains. We interpret this discrepancy as being due to the
fact that the MC simulations were carried out with $\Delta = 1$ and
$T/k_B=1$.  For this value of $\Delta$ only very large domains can
exist in the ground state.  Therefore, the broad range of analyzed
domains in \cite{cambier} results obviously from thermal fluctuations.
For larger fields and lower temperature we would expect agreement of
these two methods.

Note that all exponents found for the fractal behavior of the domains
of the RFIM as well as the DAFF are very close to those for the
lattice animals of the percolation problem \cite{stauffer,parisi}.
This is plausible since as mentioned before the domains of the RFIM in
the limit of high random fields have the same distribution and
structure like the clusters from the percolation problem. As we
discussed in the previous subsection systematic inaccuracies are not
reflected by the error-bars and systematic error can be due to a
slight curvature of the data.

Our results deviate once more from the assumption of Imry-Ma-type
random field fluctuations \cite{imry} for compact domain
structures with $\theta = 1/2$ and $b_{in/out} \sim L^{1/2}$. However,
no $\Delta$-dependence for our exponents can be determined as
indication for a crossover to Imry-Ma behavior within the range of
fields that we investigated.

\section{CONCLUSIONS}
We investigated the distribution and structure of the domains of the
ground state of DAFF and RFIM in two and three dimensions. Since
except for the Gaussian RFIM there is always a lower cut-off for the
minimum domain size one is restricted to rather large disorder due to
finite size effects, i. e. high dilution and fields of the order of
the spin-spin coupling constant.

The domain-size distribution for the DAFF can be well described by a
power law with a field dependent exponential cut-off for a broad
range of the applied fields in 2d and for strong fields ($B>3$) in
3d. For the RFIM with increasing random field strength there is a
continuous transition to the cluster-size distribution of the
percolation problem which also follows a power law with an exponential
cut-off for both Gaussian- and bimodal distribution of the
random fields. For the latter a non trivial minimum domain-size is
found, that can be estimated based on energetic considerations.

Investigating structural properties we have found that the domains are
fractal for the DAFF ($d=2,3$) as well as for the RFIM ($d=2$) with
the same fractal exponents.  Surprisingly, the domain state of the RFIM
in 3d is found to consist mainly ($> 97\%$) of two infinite
interpenetrating domains of opposite phase in the whole range of
random fields for which the long range order is broken. In this sense,
the phase diagram of the RFIM at $T=0$ consists only of two regions, a
one-domain state (long-range order below the critical field) and this
two-domain state.

While for the RFIM the results for the structural scaling exponents
differ essentially from earlier findings at finite temperature
carried out using MC simulations, our results are in good agreement
with Monte Carlo simulations for the DAFF both in two and three
dimensions.

The magnetisation of the domains and the distribution of random fields
within the domains, respectively, strongly deviates from the
assumptions of the Imry-Ma argument which however can still thought to
be correct in the limit of small disorder.  A more detailed
investigation of the structure of domains for weak random fields is
desirable.  Unfortunately, for this the numerics exceeds current
capacities.

\acknowledgements The authors would like to thank M.~Staats for 
technical support.

\newpage

\narrowtext
\begin{table}
  \begin{center}
    \begin{minipage}{2.2cm}
      \begin{tabular}{c||c}
        $B$ & $\delta$        \\ \hline \hline
        0.6 & $1.76 \pm 0.06$ \\ \hline
        0.9 & $1.76 \pm 0.06$ \\ \hline
        1.2 & $1.68 \pm 0.04$ \\ \hline
        1.4 & $1.61 \pm 0.06$ \\ \hline
        1.5 & $1.58 \pm 0.05$ \\ \hline
        2.2 & $1.44 \pm 0.03$ \\ \hline
        2.5 & $1.42 \pm 0.06$ \\ \hline
        2.8 & $1.37 \pm 0.04$ \\ \hline
        3.5 & $1.32 \pm 0.05$ \\
      \end{tabular}
    \end{minipage}
  \end{center}
  \caption {\em Field dependence for the exponent $\delta$ of the
    domain-size distribution for the 2d DAFF}
  \label{deltatable}
\end{table}

\begin{table}
  \begin{center}
    \begin{tabular}{c||c|c||c|c}  
      & \multicolumn{2}{c||} {\bf MC Simulation}
      & \multicolumn{2}{c}{\bf Exact Ground States}\\ \hline
      $d$      & 2              & 3                & 2               & 3 \\
      \hline \hline
      $\theta$ & $1.00\pm 0.01$ & $0.996\pm 0.001$ & $0.97\pm 0.01$ & $0.99\pm 0.01$\\ \hline
      $\sigma$ & $0.98\pm 0.01$ & $0.995\pm 0.001$ & $0.96\pm 0.02$ & $0.98\pm 0.03$\\ \hline
      $D_{vf}$ & $1.001\pm 0.002$ & $1.000\pm 0.001$ & $1.00\pm 0.01$ & $1.00\pm 0.01$\\ \hline
      $D_{v}$  & $1.56\pm 0.03$ & $2.0\pm 0.1$       & $1.64\pm 0.06$ & $2.14\pm 0.08$\\
    \end{tabular}
  \end{center}
  \caption{\em Structural exponents for the DAFF}
  \label{daffstructuretable}
\end{table}

\begin{table}
  \begin{center}
    \begin{tabular}{c||c|c||c|c}
      &  \multicolumn{2}{c||} {\bf MC Simulation}
      & \multicolumn{2}{c}{\bf Exact Ground States}\\ \hline
      & $\pm \Delta$   & Gaussian & $\pm \Delta$   & Gaussian \\ \hline \hline
      $D_v$   &       -        &    -     & $1.65\pm 0.08$ & $1.67 \pm 0.08$\\ \hline
      $D_{vf}$& $0.59 \pm 0.04$&    -     & $0.98\pm 0.05$ & $0.97\pm 0.05$ \\ \hline
      $\theta$& $0.66 \pm 0.04$&    -     & $1.00\pm 0.01$ & $1.00\pm 0.01$ \\
    \end{tabular}
  \end{center}
  \caption{\em Structural exponents for the 2d RFIM;
    $\Delta = 4.01$ for $\pm \Delta$- and $\Delta=3.01$ for
    Gaussian-distributed random fields}
  \label{rfimstructuretable}
\end{table}

\end{document}